\documentclass{cjaa}                   

\usepackage{graphicx}                   
\input{epsf.sty}                        
\input{psfig.sty}                       

\begin{document}

   \title{Binary Stellar Population Synthesis Study of Elliptical Galaxies
}

   \volnopage{Vol.0 (200x) No.0, 000--000}      
   \setcounter{page}{1}          

   \author{Zhongmu Li
      \inst{1,2}\mailto{}
   \and Fenghui Zhang
      \inst{1}
   \and Zhanwen Han
      \inst{1}
      }
   \offprints{Z.-M. Li}                   

   \institute{National Astronomical Observatories/Yunnan
Observatory, the Chinese Academy of Sciences, Kunming, 650011,
China\\
             \email{zhongmu.li@gmail.com}
        \and
             Graduate School of the Chinese Academy of
Sciences\\
          }

   \date{Received~~2001 month day; accepted~~2001~~month day}

   \abstract{
We determine relative stellar ages and metallicities mainly for
about 80 elliptical galaxies in low and high density environments
via the latest binary stellar population (BSP) synthesis model and
test a latest hierarchical formation model of elliptical galaxies
which adopted the new $\Lambda$CDM cosmology for the first time.
The stellar ages and metallicities of galaxies are estimated from
two high-quality published spectra line indices, i.e. H$\beta$ and
[MgFe]. The results show that elliptical galaxies have stellar
populations older than 3.9\, Gyr and more metal rich than 0.02.
Most of our results are in agreement with predictions of the
model: First, elliptical galaxies in denser environment are redder
and have older populations than field galaxies. Second, elliptical
galaxies with more massive stellar components are redder while
have older and more metal rich populations than less massive ones.
Third, the most massive galaxies are shown to have the oldest and
most metal rich stars. However, some of our results are found to
be different with predictions of the galaxy formation model, i.e.
the metallicity distributions of low- and high-density elliptical
galaxies and the relations relating to cluster-centric distance.
   \keywords{galaxies: stellar content --- galaxies: formation
   --- galaxies: elliptical and lenticular, cD }
   }

   \authorrunning{Z. M. Li, F. H., Zhang \& Z. W., Han}            
   \titlerunning{Binary population synthesis study of elliptical galaxies }  

   \maketitle

%
%
\section{Introduction}           
\label{sect:intro}
Now it is a golden era to study galaxy formation and evolution.
Fortunately, elliptical galaxies supply us a good chance to carry
out this work because they seem to be homogeneous stellar systems
that have uniformly old and red populations. Besides, elliptical
galaxies have negligible amounts of gas and have very little star
formation. Therefore, it is convenient to study galaxy formation
via ellipticals first. After the significant development of
cosmology (e.g. Peebles 1980), the image that galaxies formed in a
universe dominated by dark matter was widely accepted. But people
are still arguing about the mechanism of elliptical galaxy
formation. Recently, there are mainly two arguing pictures of
elliptical galaxies' formation. On the one hand, some people
suggest that elliptical galaxies formed in a single intense burst
of star formation at high redshifts and then their stellar
populations passively evolved to the present day. This
``monolithic'' scenario can explain the dense cores, metallicity
gradients (Kormendy 1987; Thomsen \& Baum 1989; Kormendy \&
Djorgovski 1989) and fundamental scaling relations such as the
colour-magnitude relation and the fundamental plane of elliptical
galaxies (Kodama et al. 1998; van Dokkum \& Stanford 2003), but it
cannot explain different metallicity levels of halo stars and the
big age range of globular clusters. On the other hand, based on
evidence of strong gravitational interactions and mergers between
disk galaxies, Toomre \& Toomre et al. (1972) pointed out that
elliptical galaxies are possibly formed by the merging of smaller
galaxies. It is very the so-called ``hierarchical'' scenario of
galaxy formation.

In recent years, the hierarchical picture is thought as the most
possible mechanism of galaxies and was deeply simulated, e.g.
Kauffmann et al. (1993, 1996, 1998) and Durham group (Baugh et al.
1996, Baugh et al. 1998, Cole et al. 2000). In these studies, some
exciting results are presented, e.g. the star formation histories
of galaxies (see e.g. Baugh et al. 1998). But on the observational
side, studies showed some different trends: Firstly, it is found
that a significant fraction of early-type galaxies have recent
star formation (Barger et al. 1996). Secondly, it is also found
that only a small fraction of mass is involved in the interaction
and merger of galaxies. And thirdly, some related issues, e.g. the
super-solar [$\alpha$/Fe] ratio of massive ellipticals, which
suggests these galaxies formed on relatively short time-scales and
have an initial mass function that is skewed towards massive
stars, have brought forward. It seems that those early models of
the hierarchical formation of elliptical galaxies are difficult to
explain and reproduce these observed trends (Thomas 1999). In
2006, De Lucia et al. (2006) brought forward a new hierarchical
model of the formation of elliptical galaxies. This model,
adopting the new $\Lambda$CDM cosmology and high-resolution
simulation, tried to explain the ``anti-hierarchical'' behavior of
star formation histories of elliptical galaxy population and
presented some new predictions. Therefore, it is more valuable to
study the formation of elliptical galaxies based on this kind new
models now.

To study galaxy formation, stellar population synthesis has become
a very useful and popular technique in these years because
different galaxy formation models usually predict different star
formation histories.  A series of detailed studies of stellar
populations of galaxies in both observational side and
semi-analytic side have been carried out in recent years (e.g.
Trager et al. 2000a, b; Terlevich \& A. Forbes D. 2002; van Zee et
al. 2004). However, all these works used the single stellar
population (SSP) synthesis models (e.g. Vazdekis et al. 1996,
1997; Vazdekis 1999; Worthey 1994; Bruzual \& Charlot et al. 2003)
as the binary stellar population (BSP) synthesis model was not
available. But as pointed out by Zhang et al. (2005a, b), binary
interaction plays an important role in the study of evolutionary
population synthesis. In their work, some different results from
SSP models were shown (see Zhang 2005a in more detail). Thus we
are now asking the question that how hierarchical formation model
of elliptical galaxies is supported if we take the binary
interaction in stellar population synthesis into account. We plan
to find some answers by making use of the BSP model of Zhang et
al. (2005b) and the hierarchical formation model of De Lucia et
al. (2006) in this paper. But we here do not intend to investigate
the effects of binary interaction, which is very the subject of
another paper. The structure of the paper is as follows. In Sect.
2 we introduce our galaxy sample and the BSP model. In Sect. 3 we
give a brief description of the determination of stellar ages and
metallicities and then show the main results. In Sect. 4 we test
the latest hierarchical formation model of elliptical galaxies and
finally we give our discussion and conclusion in Sect. 5.


\section{Our data sample and the BSP model}
\label{sect:Obs}
\subsection{ The galaxy sample}
We mainly define a sample by selecting all normal ellicptical
galaxies in the sample of Thomas et al. (2005). As a result, 71
normal elliptical galaxies are included while 51 S0 and 2 cD
galaxies are excluded by our sample. Then the $B-V$ colors and
B-band absolute magnitudes of these galaxies are supplemented from
the HyperLeda database (http://www.brera.mi.astro.it/hypercat/) if
it is possible. In the sample, 42 elliptical galaxies reside in
low-density and 29 in high-density environments. The galaxies in
low-density environment include all galaxies that do not reside in
high-density environment. In fact, these data are very good for
estimating stellar ages and metallicities because they were
selected from some creditable sources (Gonz$\acute{a}$lez 1993;
Mehlert et al. 2000, 2003; Beuing et al. 2002; Lauberts \&
Valentijn 1989) and reobserved by Thomas et al. if necessary (19
objects were reobserved). In particular, because the
absorption-line strengths of galaxies are measured as functions of
galaxy radius in the sources, the central indices measured within
r$_{e}$/10 (where r$_{e}$ is the effective radius) are adopted, so
that the analysis does not suffer from aperture effects. In this
work, we use the reliable Lick indices H$\beta$, Mgb, and
$<$Fe$>$=0.5$\times$(Fe5270+Fe5335) directly. According to Thomas
et al., the medians of the 1$\sigma$ errors in H$\beta$, Mgb,
Fe5270 and Fe5335 are 0.06,0.06,0.07 and 0.08, respectively. It is
also another advantage to take these data because these elliptical
galaxies span a large range in central velocity dispersion 0
$\leq$ $\sigma$$_0$/(km s$^{-1}$) $\leq$ 340, which is very
convenient to study the relations between stellar specialities and
stellar mass following the result of Thomas et al. (2005). The
detailed data of our sample galaxies are shown in Table 1. In the
table, the galaxy name, velocity dispersion, H$\beta$, Mgb,
$<$Fe$>$, M$_B$, $B-V$, environment and the observational
uncertainties of three line indices are shown. Besides, we also
select 11 elliptical galaxies in the Fornax cluster from
Kuntschner (2000), but they are only used for testing the
predictions relating to cluster-centric distance.

\begin{table}[]
\caption[]{The data for low- and high-density ellipticals. In the
table, `$\sigma$$_{\rm 0}$' means the velocity dispersion and `E'
means the environment. `L' and `H' denotes low- and high-density
environments, respectively. All line indices are within r$_e$/10
aperture.} \label{Tab:1}
\begin{center}\begin{tabular}{lrrrrrrrrrr}
\hline\noalign{\smallskip}
\multicolumn {1} {l} {Name}& \multicolumn {1} {c} {$\sigma$$_{\rm
0}$} &\multicolumn {1} {c} {H$\beta$}&\multicolumn {1} {c}
{$\delta$H$\beta$}&\multicolumn {1} {c} {Mgb}&\multicolumn {1} {c}
{$\delta$Mgb}& \multicolumn {1} {c} {$<$Fe$>$} &\multicolumn {1}
{c} {$\delta$$<$Fe$>$} & \multicolumn {1} {c} {$M_B$}
&\multicolumn {1} {c} {$B-V$}&\multicolumn {1} {c} {E}\\

&\multicolumn {1} {c}{[km s$^{-1}$]} &\multicolumn {1} {c}{[\AA]} &\multicolumn {1} {c}{[\AA]}
&\multicolumn {1} {c}{[\AA]} &\multicolumn {1} {c}{[\AA]}&\multicolumn {1} {c}{[\AA]}
&\multicolumn {1} {c}{[\AA]} &\multicolumn {1} {c}{[mag]}&\multicolumn {1} {c}{[mag]}\\
\hline
NGC 0221   & 72.1 &2.31 &0.05 &2.96 &0.03 &2.75 &0.03 &-17.424 &0.800 &L \\
NGC 0315   &321.0 &1.74 &0.06 &4.84 &0.05 &2.88 &0.05 &-22.472 &0.929 &L \\
NGC 0507   &262.2 &1.73 &0.09 &4.52 &0.11 &2.78 &0.10 &-22.121 &0.888 &L \\
NGC 0547   &235.6 &1.58 &0.07 &5.02 &0.05 &2.82 &0.05 &-21.663 &      &L \\
NGC 0636   &160.3 &1.89 &0.04 &4.20 &0.04 &3.03 &0.04 &-19.798 &0.908 &L \\
NGC 0720   &238.6 &1.77 &0.12 &5.17 &0.11 &2.87 &0.09 &-20.786 &0.948 &L \\
NGC 0821   &188.7 &1.66 &0.04 &4.53 &0.04 &2.95 &0.04 &-20.753 &0.865 &L \\
NGC 1453   &286.5 &1.60 &0.06 &4.95 &0.05 &2.98 &0.05 &-21.613 &0.911 &L \\
NGC 1600   &314.8 &1.55 &0.07 &5.13 &0.06 &3.06 &0.06 &-22.419 &0.923 &L \\
NGC 1700   &227.3 &2.11 &0.05 &4.15 &0.04 &3.00 &0.04 &-21.903 &0.890 &L \\
NGC 2300   &251.8 &1.68 &0.06 &4.98 &0.05 &2.97 &0.05 &-20.754 &0.966 &L \\
NGC 2778   &154.4 &1.77 &0.08 &4.70 &0.06 &2.85 &0.05 &-19.206 &0.889 &L \\
NGC 3377   &107.6 &2.09 &0.05 &3.99 &0.03 &2.61 &0.03 &-19.169 &0.820 &L \\
NGC 3379   &203.2 &1.62 &0.05 &4.78 &0.03 &2.86 &0.03 &-20.608 &0.927 &L \\
NGC 3608   &177.7 &1.69 &0.06 &4.61 &0.04 &2.94 &0.04 &-19.733 &0.909 &L \\
NGC 3818   &173.2 &1.71 &0.08 &4.88 &0.07 &2.97 &0.06 &-19.400 &0.908 &L \\
NGC 4278   &232.5 &1.56 &0.05 &4.92 &0.04 &2.68 &0.04 &-19.359 &0.895 &L \\
NGC 5638   &154.2 &1.65 &0.04 &4.64 &0.04 &2.84 &0.04 &-19.974 &0.892 &L \\
NGC 5812   &200.3 &1.70 &0.04 &4.81 &0.04 &3.06 &0.04 &-20.450 &0.927 &L \\
NGC 5813   &204.8 &1.42 &0.07 &4.65 &0.05 &2.67 &0.05 &-21.113 &0.916 &L \\
NGC 5831   &160.5 &2.00 &0.05 &4.38 &0.04 &3.05 &0.03 &-19.813 &0.897 &L \\
NGC 6127   &238.9 &1.50 &0.05 &4.96 &0.06 &2.85 &0.05 &-21.352 &0.944 &L \\
NGC 6702   &173.8 &2.46 &0.06 &3.80 &0.04 &3.00 &0.04 &-21.613 &0.839 &L \\
NGC 7052   &273.8 &1.48 &0.07 &5.02 &0.06 &2.84 &0.05 &-21.199 &      &L \\
NGC 7454   &106.5 &2.15 &0.06 &3.27 &0.05 &2.48 &0.04 &-19.930 &0.866 &L \\
NGC 7785   &239.6 &1.63 &0.06 &4.60 &0.04 &2.91 &0.04 &-21.375 &0.949 &L \\
ESO 107-04 &147.0 &2.24 &0.25 &3.63 &0.16 &2.97 &0.09 &-20.386 &0.849 &L \\
ESO 148-17 &134.5 &2.26 &0.52 &3.49 &0.32 &2.58 &0.20 &-19.865 &0.875 &L \\
IC 4797    &220.6 &1.92 &0.26 &4.52 &0.18 &2.75 &0.10 &-20.876 &0.908 &L \\
NGC 0312   &254.8 &1.83 &0.09 &4.56 &0.08 &2.48 &0.05 &-21.937 &0.929 &L \\
NGC 0596   &161.8 &2.12 &0.05 &3.95 &0.04 &2.81 &0.03 &-20.424 &0.845 &L \\
NGC 0636   &178.5 &1.86 &0.26 &4.38 &0.17 &2.83 &0.09 &-19.798 &0.908 &L \\
NGC 1052   &202.6 &1.22 &0.04 &5.53 &0.03 &2.77 &0.02 &-20.139 &0.900 &L \\
NGC 1395   &250.0 &1.62 &0.05 &5.21 &0.04 &2.93 &0.03 &-21.211 &0.921 &L \\
NGC 1407   &259.7 &1.67 &0.07 &4.88 &0.06 &2.85 &0.03 &-21.432 &0.946 &L \\
NGC 1549   &203.3 &1.79 &0.03 &4.39 &0.03 &2.88 &0.02 &-19.981 &0.906 &L \\
NGC 2434   &180.4 &1.87 &0.13 &3.72 &0.10 &2.87 &0.07 &-19.828 &0.818 &L \\
NGC 2986   &282.2 &1.48 &0.06 &4.97 &0.05 &2.92 &0.03 &-21.064 &0.891 &L \\
NGC 3078   &268.1 &1.12 &0.09 &5.20 &0.07 &3.16 &0.04 &-20.893 &0.916 &L \\
NGC 3923   &267.9 &1.87 &0.08 &5.12 &0.07 &3.07 &0.04 &-21.151 &0.906 &L \\
NGC 5791   &271.8 &1.60 &0.19 &5.06 &0.15 &3.30 &0.10 &-21.123 &0.89  &L \\
NGC 5903   &209.2 &1.68 &0.10 &4.44 &0.08 &2.90 &0.05 &-21.220 &0.839 &L \\
NGC 4261   &288.3 &1.34 &0.06 &5.11 &0.04 &3.01 &0.04 &-21.299 &0.952 &H \\
NGC 4374   &282.1 &1.51 &0.04 &4.78 &0.03 &2.82 &0.03 &-20.888 &0.931 &H \\
NGC 4472   &279.2 &1.62 &0.06 &4.85 &0.06 &2.91 &0.05 &-21.785 &0.928 &H \\
NGC 4478   &127.7 &1.84 &0.06 &4.33 &0.06 &2.94 &0.05 &-19.564 &0.873 &H \\
\noalign{\smallskip}\hline
\end{tabular}\end{center}
\end{table}

\addtocounter{table} {-1}
\begin{table}[]
\caption[]{-- Continued} \label{Tab:1}
\begin{center}\begin{tabular}{lrrrrrrrrrr}
\hline\noalign{\smallskip}
\multicolumn {1} {l} {Name}& \multicolumn {1} {c} {$\sigma$$_{\rm
0}$} &\multicolumn {1} {c} {H$\beta$}&\multicolumn {1} {c}
{$\delta$H$\beta$}&\multicolumn {1} {c} {Mgb}&\multicolumn {1} {c}
{$\delta$Mgb}& \multicolumn {1} {c} {$<$Fe$>$} &\multicolumn {1}
{c} {$\delta$$<$Fe$>$} & \multicolumn {1} {c} {$M_B$}
&\multicolumn {1} {c} {$B-V$}&\multicolumn {1} {c} {E}\\

&\multicolumn {1} {c}{[km s$^{-1}$]} &\multicolumn {1} {c}{[\AA]}
&\multicolumn {1} {c}{[\AA]} &\multicolumn {1} {c}{[\AA]}
&\multicolumn {1} {c}{[\AA]}&\multicolumn {1} {c}{[\AA]}
&\multicolumn {1} {c}{[\AA]} &\multicolumn {1} {c}{[mag]}&\multicolumn {1} {c}{[mag]}\\
\hline
NGC 4489   & 47.2 &2.39 &0.07 &3.21 &0.06 &2.66 &0.05 &-18.189 &0.804 &H \\
NGC 4552   &251.8 &1.47 &0.05 &5.15 &0.03 &2.99 &0.03 &-20.798 &0.936 &H \\
NGC 4697   &162.4 &1.75 &0.07 &4.08 &0.05 &2.77 &0.04 &-21.239 &0.869 &H \\
NGC 7562   &248.0 &1.69 &0.05 &4.54 &0.04 &2.87 &0.04 &-21.416 &0.938 &H \\
NGC 7619   &300.3 &1.36 &0.04 &5.06 &0.04 &3.06 &0.04 &-21.973 &0.925 &H \\
NGC 7626   &253.1 &1.46 &0.05 &5.05 &0.04 &2.83 &0.04 &-21.673 &0.947 &H \\
NGC 4839   &275.5 &1.42 &0.04 &4.92 &0.04 &2.75 &0.04 &-22.263 &0.879 &H \\
NGC 4841A  &263.9 &1.53 &0.05 &4.51 &0.05 &2.89 &0.04 &-21.380 &      &H \\
NGC 4926   &273.3 &1.50 &0.06 &5.17 &0.06 &2.50 &0.05 &-21.443 &0.954 &H \\
IC 4051    &258.7 &1.42 &0.06 &5.34 &0.07 &2.75 &0.05 &-20.204 &0.933 &H \\
NGC 4860   &280.5 &1.39 &0.06 &5.39 &0.07 &2.85 &0.05 &-20.948 &0.973 &H \\
NGC 4923   &186.0 &1.70 &0.05 &4.43 &0.05 &2.69 &0.04 &-19.983 &0.888 &H \\
NGC 4840   &216.6 &1.63 &0.07 &4.94 &0.07 &2.91 &0.06 &-20.131 &0.954 &H \\
NGC 4869   &188.1 &1.40 &0.05 &4.83 &0.05 &2.90 &0.04 &-20.797 &0.934 &H \\
NGC 4908   &192.4 &1.58 &0.09 &4.58 &0.09 &2.65 &0.07 &-21.075 &0.936 &H \\
IC 4045    &167.3 &1.46 &0.06 &4.70 &0.07 &2.77 &0.05 &-20.282 &0.943 &H \\
NGC 4850   &155.8 &1.57 &0.06 &4.39 &0.06 &2.58 &0.05 &-19.601 &0.956 &H \\
NGC 4872   &171.7 &2.05 &0.05 &4.05 &0.06 &2.82 &0.04 &-20.893 &0.874 &H \\
NGC 4957   &208.4 &1.76 &0.03 &4.53 &0.03 &2.93 &0.02 &-21.177 &0.925 &H \\
NGC 4952   &252.6 &1.71 &0.03 &4.76 &0.03 &2.69 &0.02 &-21.203 &      &H \\
GMP 1990   &208.9 &1.40 &0.04 &4.78 &0.04 &2.50 &0.03 &        &      &H \\
NGC 4827   &243.7 &1.53 &0.03 &4.89 &0.03 &2.80 &0.02 &-21.495 &0.904 &H \\
NGC 4807   &178.5 &1.81 &0.06 &4.39 &0.06 &2.78 &0.05 &-20.703 &0.919 &H \\
ESO 185-54 &277.2 &1.57 &0.06 &5.11 &0.05 &3.07 &0.03 &-21.861 &      &H \\
NGC 3224   &155.8 &2.31 &0.14 &3.91 &0.12 &2.92 &0.08 &-20.508 &0.828 &H \\

\noalign{\smallskip}\hline
\end{tabular}\end{center}
\end{table}

\subsection{ The BSP model}
In this work, we translate central line indices of galaxies into
stellar ages and metallicities via the BSP model of Zhang et al.
(2005b). This model supplied us with high-resolution (0.3\, \AA)
absorption-lines defined by the Lick Observatory Image Dissector
Scanner (Lick/IDS) system for an extensive set of instantaneous
burst binary stellar populations with binary interactions. In
particular, its stellar populations span an age range 1--15\, Gyr
and a metallicity range 0.004--0.03.

\section{Stellar ages and metallicities of elliptical galaxies}
\label{sect:data}
To determine BSP-equivalent stellar ages and metallicities of
elliptical galaxies, we use H$\beta$ and [MgFe] (Gonz$\acute
{a}$lez 1993) indices in this work. The latter can be calculated
by $\rm \sqrt{Mgb\times0.5\times(Fe5270+Fe5335)}$. Fig. 1 displays
the H$\beta$ and [MgFe] indices of the 71 elliptical galaxies
overlaid onto the theoretical calibration. The open and filled
circles represent ellipticals in low- and high-density
environments respectively and the error bars show the
observational uncertainties of two indices.

%
\begin{figure}
   \vspace{2mm}
   \begin{center}
   \hspace{3mm}\psfig{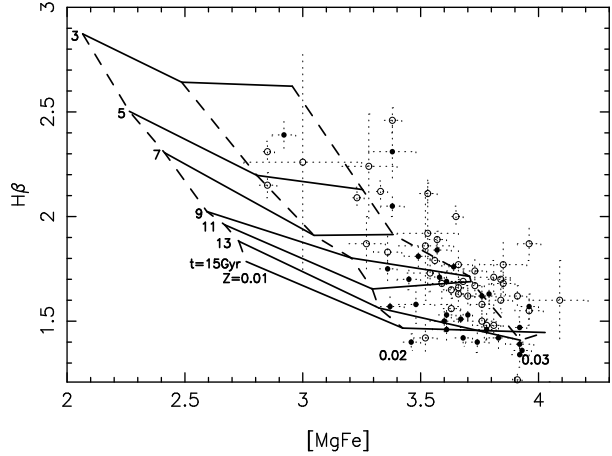}
   \parbox{180mm}{{\vspace{2mm} }}
   \caption{ Line-strength indices of our sample elliptical galaxies in the BSP model
   in the central r$_e$/10 aperture. Solid lines represent constant age (isochrones)
   and dashed lines constant metallicity (isofers). Open and filled circles represent low- and
   high-density ellipticals, respectively. Error bars show the observational
   uncertainties of two indices.}
   \label{Fig:lightcurve-ADAri}
   \end{center}
\end{figure}

The BSP-equivalent stellar age, metallicity of each elliptical
galaxy is determined by choosing the best-fitting (t, Z) in a grid
of stellar age (t) and metallicity (Z). The grid is elaborate
enough and created by interpolating the BSP models at intervals
$\Delta$t = 0.1\, Gyr and $\Delta$Z = 0.0001. To find the
best-fitting (t, Z), we employ the maximum-likelihood fitting
method. In detail, we obtain the best-fitting age and metallicity
of each galaxy by minimizing the function:
\begin{equation}
     \rm \chi^{2}(t_{\it i},Z_{\it i})=(H\beta_{\it i}-H\beta_{o})^{2}
     +({[MgFe]}_{\it i}-{[MgFe]}_{o})^{2},
\end{equation}
where H$\beta$$_{i}$ and [MgFe]$_{i}$ are the H$\beta$ and [MgFe]
indices corresponding to the \emph {i}th pair of stellar age and
metallicity in the BSP model, while H$\beta$$\rm _o$ and
[MgFe]$\rm _o$ are two observational indices. Moreover, we
calculate the associated uncertainties of the best-fitting stellar
age and metallicity for each galaxy by searching best-fitting (t,
Z)s for [H$\beta$$\rm _o$-error, [MgFe]$\rm _o$], [H$\beta$$\rm
_o$+error, [MgFe]$\rm _o$], [H$\beta$$\rm _o$, [MgFe]$\rm
_o$-error] and [H$\beta$$\rm _o$, [MgFe]$\rm _o$+error]
respectively and then taking their deviations from the
best-fitting (t, Z) derived from [H$\beta$$\rm _o$, [MgFe]$\rm
_o$]. Although the four pairs (t, Z) derived by searching do not
describe perfectly well the total range of possible age and
metallicity values inside the 1$\sigma$ error ellipse, the maximum
deviations of stellar age and metallicity can provide us with a
sufficient sampling of the uncertainties involved when transposing
errors from the H$\beta$-[MgFe] plane to ages or metallicities
(Denicol\'{o} et al. 2005). Therefore, in this work, we take the
maximum deviation as the associated 1$\sigma$ uncertainty for
stellar age and metallicity. Here we show the stellar ages,
metallicities and their 1$\sigma$ uncertainties of the main sample
galaxies in Table 2. The stellar ages of these ellipticals are
within the range from 3.9 to older than 15\, Gyr and the stellar
metallicities span over the range of 0.02 to richer than 0.03. It
seems that these ages do not vary as widely as Trager et al.
(2000a) whose result is 1.5 -- 18\, Gyr. This should result from
the different stellar population synthesis model adopted in this
paper. It is found that about 78\% elliptical galaxies have
stellar populations older than 8\, Gyr. The average stellar age of
elliptical galaxies is 10.37\, Gyr while the average of
metallicity is 0.0277. The average 1$\sigma$ uncertainties of them
are 1.58\, Gyr and 0.0015, respectively.

\begin{table}[]
\caption[]{Stellar ages, metallicities and associated 1$\sigma$
uncertainties of 71 sample elliptical galaxies. The stellar ages
and their uncertainties are in Gyr. } \label{Tab:2}
\begin{center}\begin{tabular}{lrrlrr}
\hline\noalign{\smallskip}
\multicolumn {1} {l} {Name}& \multicolumn {1} {c} {Age}
&\multicolumn {1} {c} {Z}&\multicolumn {1} {l} {Name}&
\multicolumn {1} {c} {Age} &\multicolumn {1} {c} {Z}\\
\hline\noalign{\smallskip}
NGC 0221    & 4.3       $\pm$ 0.3 &0.0230     $\pm$ 0.0008        &NGC 2434   & 8.3       $\pm$ 5.9          &0.0234     $\pm$ 0.0041 \\
NGC 0315    & 9.0       $\pm$ 2.1 &\multicolumn{1}{l}{$\geq$0.03} &NGC 2986   &12.6       $\pm$ 2.0          &0.0284     $\pm$ 0.0011 \\
NGC 0507    & 9.2       $\pm$ 2.3 &0.0268     $\pm$ 0.0023        &NGC 3078   &13.9       $\pm$ 0.5          &\multicolumn{1}{l}{$\geq$0.03}\\
NGC 0547    &11.8       $\pm$ 2.7 &0.0286     $\pm$ 0.0025        &NGC 3923   & 9.0       $\pm$ 2.5          &\multicolumn{1}{l}{$\geq$0.03}\\
NGC 0636    & 7.8       $\pm$ 0.2 &\multicolumn{1}{l}{$\geq$0.03} &NGC 5791   &$\geq$15   $\pm$ 3.2          &\multicolumn{1}{l}{$\geq$0.03}\\
NGC 0720    &11.2       $\pm$ 2.2 &\multicolumn{1}{l}{$\geq$0.03} &NGC 5903   &10.3       $\pm$ 4.1          &0.0276     $\pm$ 0.0040 \\
NGC 0821    &11.2       $\pm$ 1.7 &0.0280     $\pm$ 0.0012        &NGC 4261   &13.0       $\pm$ 0.3          &$\geq$0.03 $\pm$ 0.0004 \\
NGC 1453    &11.7       $\pm$ 0.3 &$\geq$0.03 $\pm$ 0.0006        &NGC 4374   &13.3       $\pm$ 0.6          &0.0256     $\pm$ 0.0004 \\
NGC 1600    &14.0       $\pm$ 2.0 &0.0286     $\pm$ 0.0014        &NGC 4472   &11.4       $\pm$ 2.6          &0.0296     $\pm$ 0.0032 \\
NGC 1700    & 6.0       $\pm$ 0.1 &\multicolumn{1}{l}{$\geq$0.03} &NGC 4478   & 7.9       $\pm$ 0.7          &$\geq$0.03 $\pm$ 0.0017 \\
NGC 2300    &11.4       $\pm$ 0.2 &\multicolumn{1}{l}{$\geq$0.03} &NGC 4489   & 3.9       $\pm$ 0.3          &0.0254     $\pm$ 0.0012 \\
NGC 2778    & 8.6       $\pm$ 1.0 &$\geq$0.03 $\pm$ 0.0010        &NGC 4552   &14.3       $\pm$ 1.8          &0.0285     $\pm$ 0.0014 \\
NGC 3377    & 5.4       $\pm$ 0.4 &0.0283     $\pm$ 0.0011        &NGC 4697   & 9.3       $\pm$ 1.1          &0.0229     $\pm$ 0.0014 \\
NGC 3379    &11.5       $\pm$ 2.7 &0.0281     $\pm$ 0.0027        &NGC 7562   & 9.7       $\pm$ 1.7          &0.0283     $\pm$ 0.0017 \\
NGC 3608    & 9.6       $\pm$ 1.8 &0.0298     $\pm$ 0.0017        &NGC 7619   &13.0       $\pm$ 0.4          &$\geq$0.03 $\pm$ 0.0003 \\
NGC 3818    &11.2       $\pm$ 2.2 &\multicolumn{1}{l}{$\geq$0.03} &NGC 7626   &13.2       $\pm$ 0.8          &0.0274     $\pm$ 0.0007 \\
NGC 4278    &12.3       $\pm$ 2.2 &0.0258     $\pm$ 0.0023        &NGC 4839   &\multicolumn{1}{l}{$\geq$15.0}&0.0242     $\pm$ 0.0007 \\
NGC 5638    &11.3       $\pm$ 1.6 &0.0272     $\pm$ 0.0015        &NGC 4841A  &12.6       $\pm$ 2.3          &0.0252     $\pm$ 0.0020 \\
NGC 5812    &11.3       $\pm$ 0.2 &\multicolumn{1}{l}{$\geq$0.03} &NGC 4926   &12.9       $\pm$ 2.1          &0.0246     $\pm$ 0.0018 \\
NGC 5813    &$\geq$15   $\pm$ 0.1   &0.0217     $\pm$ 0.0009      &IC 4051    &13.0       $\pm$ 0.6          &0.0285     $\pm$ 0.0009 \\
NGC 5831    & 8.0       $\pm$ 0.2 &\multicolumn{1}{l}{$\geq$0.03} &NGC 4860   &13.0       $\pm$ 2.0          &$\geq$0.03 $\pm$ 0.0018 \\
NGC 6127    &13.5       $\pm$ 1.1 &0.0268     $\pm$ 0.0019        &NGC 4923   & 9.8       $\pm$ 4.0          &0.0247     $\pm$ 0.0031 \\
NGC 6702    & 4.0       $\pm$ 0.0 &\multicolumn{1}{l}{$\geq$0.03} &NGC 4840   &11.4       $\pm$ 0.5          &$\geq$0.03 $\pm$ 0.0012 \\
NGC 7052    &12.7       $\pm$ 2.0 &0.0278     $\pm$ 0.0010        &NGC 4869   &13.0       $\pm$ 2.0          &0.0272     $\pm$ 0.0027 \\
NGC 7454    & 5.3       $\pm$ 0.8 &0.0200     $\pm$ 0.0021        &NGC 4908   &12.4       $\pm$ 2.5          &0.0234     $\pm$ 0.0022 \\
NGC 7785    &11.4       $\pm$ 2.6 &0.0276     $\pm$ 0.0031        &IC 4045    &$\geq$15   $\pm$ 0.3          &0.0230     $\pm$ 0.0010 \\
ESO 107-04  & 4.6       $\pm$ 1.6 &$\geq$0.03 $\pm$ 0.0022        &NGC 4850   &12.8       $\pm$ 2.1          &0.0210     $\pm$ 0.0016 \\
ESO 148-17  & 4.5       $\pm$ 8.6 &0.0255     $\pm$ 0.0109        &NGC 4872   & 5.8       $\pm$ 0.2          &\multicolumn{1}{l}{$\geq$0.03}\\
IC 4797     & 7.6       $\pm$ 6.3 &$\geq$0.03 $\pm$ 0.0067        &NGC 4957   & 8.6       $\pm$ 0.3          &0.0297     $\pm$ 0.0007 \\
NGC 0312    & 8.5       $\pm$ 1.6 &0.0246     $\pm$ 0.0048        &NGC 4952   & 9.4       $\pm$ 1.1          &0.0275     $\pm$ 0.0005 \\
NGC 0596    & 5.3       $\pm$ 0.2 &\multicolumn{1}{l}{$\geq$0.03} &GMP 1990   &\multicolumn{1}{l}{$\geq$15.0}&0.0206     $\pm$ 0.0006 \\
NGC 0636    & 7.7       $\pm$ 4.4 &0.0299     $\pm$ 0.0056        &NGC 4827   &12.4       $\pm$ 1.2          &0.0268     $\pm$ 0.0012 \\
NGC 1052    &13.0       $\pm$ 0.0 &\multicolumn{1}{l}{$\geq$0.03} &NGC 4807   & 8.4       $\pm$ 0.8          &0.0274     $\pm$ 0.0023 \\
NGC 1395    &11.8       $\pm$ 0.2 &\multicolumn{1}{l}{$\geq$0.03} &ESO 185-54 &12.0       $\pm$ 2.0          &$\geq$0.03 $\pm$ 0.0012 \\
NGC 1407    &11.1       $\pm$ 2.1 &$\geq$0.03 $\pm$ 0.0015        &NGC 3224   & 4.4       $\pm$ 0.9          &\multicolumn{1}{l}{$\geq$0.03}\\
NGC 1549    & 8.5       $\pm$ 0.4 &0.0283     $\pm$ 0.0010        \\

\noalign{\smallskip}\hline
\end{tabular}\end{center}
\end{table}

\section{The test of new hierarchical model}
\label{sect:test} The hierarchical formation of elliptical
galaxies has been simulated by many techniques, for example the
N-body simulation and semi-analytic simulation techniques.
Different models were usually carried out in the framework of a
cosmological model with critical matter density and gave different
predictions of stellar properties. By now, the cosmology used
before has been replaced by the $\Lambda$CDM scenario. In this
background, De Lucia et al. (2006) constructed a new hierarchical
formation model of elliptical galaxies based on the $\Lambda$CDM
scenario cosmology and studied how the star formation histories,
ages and metallicities of elliptical galaxies depend on
environment and on stellar mass. As a result, some special
predictions are presented by this model. Firstly, it predicted
that the populations of ellipticals in high-density environment
would be older, more metal rich and redder than those of field
ellipticals. Secondly, it predicted that the most massive
elliptical galaxies would have the oldest and most metal rich
stellar populations. Thirdly, it predicted that the stellar age,
metallicity and galaxy color would increase with increasing
stellar mass. Fourthly, the stellar mass, ages, metallicities and
colors of cluster elliptical galaxies were predicted to decrease
on average with increasing distance from the cluster center. In
addition, the model quantified the effective progenitors of
ellipticals as a function of present stellar mass and then
predicted the ``down-sizing'' or ``anti-hierarchical'' of star
formation histories of ellipticals in a $\Lambda$CDM universe. It
is an important result, because if this model is right, we will
understand the formation of elliptical galaxies much better.
Therefore, it is very necessary to test this model. Of course,
taking the binary interaction into to the test is important
because more than half of stars are binaries as we know. The
detailed tests are as follows.

\subsection{Stellar age, metallicity and galaxy color variation with environments}
A basic prediction of hierarchical galaxy formation picture is
that stellar populations of more massive galaxies are older than
those of less massive galaxies on average (e.g. Kauffmann 1996).
This is also predicted by the model of De Lucia et al. (2006).
Furthermore, De Lucia et al.'s model predicted that galaxies in
denser environment would have more metal rich and redder
populations than ellipticals. These specialties are thought to be
attributed to the fact that high density regions form from the
highest density peaks in the primordial field of density
fluctuations. Here we test these specialties with our data.

%
\begin{figure}
   \vspace{2mm}
   \begin{center}
   \hspace{3mm}\psfig{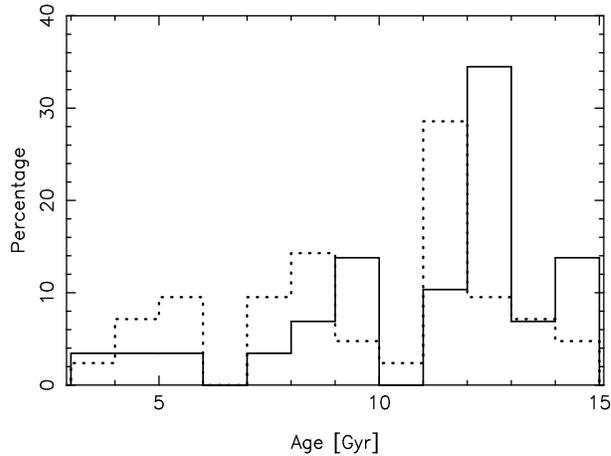}
   \parbox{180mm}{{\vspace{2mm} }}
   \caption{Stellar age distribution of low- and high-density elliptical
   galaxies. The dashed and solid lines show the distribution of low- and
   high-density ellipticals, respectively.}
   \label{Fig:lightcurve-ADAri}
   \end{center}
\end{figure}

In Fig. 2, we show the stellar age distributions of both low- and
high-density ellipticals. The dashed lines represent the stellar
age distribution of ellipticals in low-density environment while
the solid lines for high-density ellipticals. We see that the
stellar populations of high-density ellipticals are really older
than those of low-density complements. On average, stellar
populations of high-density ellipticals are 1.47\, Gyr older than
those of low-density ellipticals.

%
\begin{figure}
   \vspace{2mm}
   \begin{center}
   \hspace{3mm}\psfig{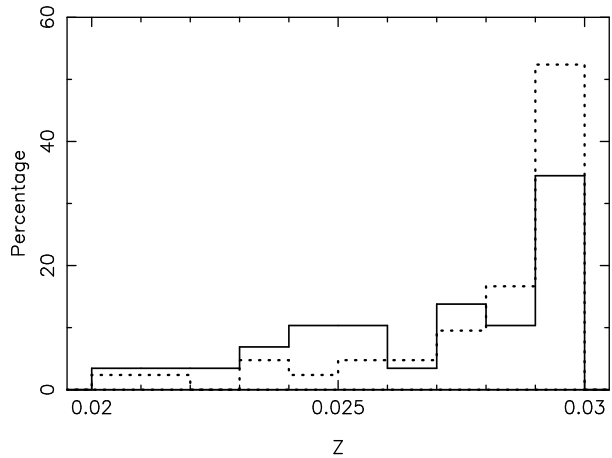}
   \parbox{180mm}{{\vspace{2mm} }}
   \caption{ Stellar metallicity distribution of low- and high-density elliptical
   galaxies. The dashed and solid lines have the same meanings as in Fig. 2. }
   \label{Fig:lightcurve-ADAri}
   \end{center}
\end{figure}

In Fig. 3, we show the stellar metallicity distributions of low
and high-density ellipticals. As we see, the plot fails to show
that stellar populations of high-density ellipticals are more
metal rich than those of low-density ellipticals. We also can see
this trend clearly from Fig. 1 that ellipticals in high-density
environment really distribute in the lower metallicity region than
those field ellipticals. In the figure, filled circles represent
ellipticals in high-density environment.

%
\begin{figure}
   \vspace{2mm}
   \begin{center}
   \hspace{3mm}\psfig{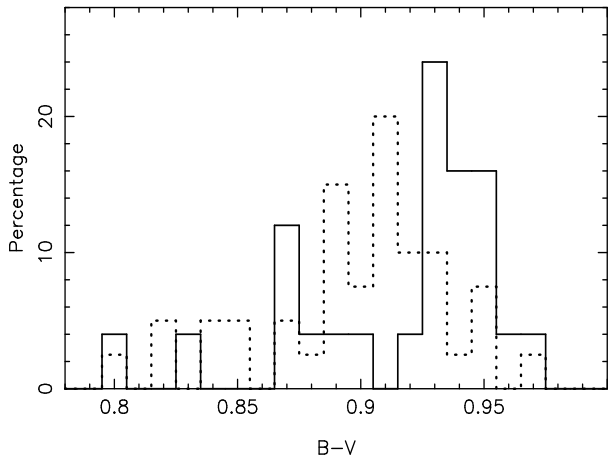}
   \parbox{180mm}{{\vspace{2mm} }}
   \caption{ $B-V$ distribution of low- and high-density elliptical
   galaxies. The dashed and solid lines have the same meanings as in Fig. 2.}
   \label{Fig:lightcurve-ADAri}
   \end{center}
\end{figure}

When we study the $B-V$ color distribution of two type elliptical
galaxies, the result is consistent with De Lucia (2006) model (see
Fig. 4 in more detail). On average, ellipticals in high-density
environment are about 0.02 mag redder than those in low-density
environment.

\subsection{Stellar age, metallicity and galaxy color variation with stellar mass}
The most important result and prediction of the De Lucia model is
that the most massive elliptical galaxies have the oldest and most
metal rich stellar populations. Besides, the model predicted that
the stellar age, metallicity and galaxy color would increase with
increasing stellar mass. We test these predictions in Figs 5, 6
and 7, respectively. Here the stellar masses of elliptical
galaxies are calculated by the fitting function suggested by
Thomas et al. (2005):
\begin{equation}
    \rm log(\it M_{\ast}/M{_\odot}) = \rm 0.63 + 4.52 log(\it \sigma_{\rm 0}/(\rm km \ s^{-1})) ,
\end{equation}
where \it{M$_\ast$} \rm is the stellar mass and $\sigma$$_{\rm 0}$
is the velocity dispersion. According to previous studies, there
is usually a relation between the mass and luminosity of
elliptical galaxies, e.g. $(M/L)$$_B$=(5.93$\pm$0.25)$h$$_{50}$
(van der Marel 1991). It means that the luminous elliptical
galaxies have the massive masses. Therefore the absolute magnitude
is usually used for a indicator of stellar mass of galaxies (e.g.
Terlevich \& A. Forbes D. 2002). In order to study the reliability
of the mass estimation presented above, we compare the stellar
masses calculated by the function with absolute B-band magnitudes
of these galaxies, which are taken from HyperLeda database
(http://www.brera.mi.astro.it/hypercat/). As a result, we find
that luminous elliptical galaxies have more massive stellar
masses. Therefore, as a whole, the stellar masses estimated by the
fitting function can express the real stellar masses of elliptical
galaxies well.

%
\begin{figure}
   \vspace{2mm}
   \begin{center}
   \hspace{3mm}\psfig{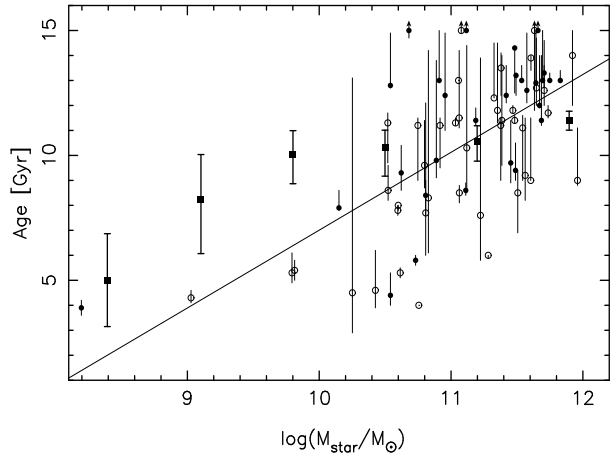}
   \parbox{180mm}{{\vspace{2mm} }}
   \caption{ Stellar age - mass relation of 71 elliptical galaxies. Filled
   squares with error bars are predictions of galaxy formation model. With arrows,
   some galaxies that have stellar populations possibly older than the maximum age of BSP model (15\, Gyr)
   are shown. The solid line is a linear least-squares fit to the data. Open and filled circles represent low- and
   high-density ellipticals, respectively.}
   \label{Fig:lightcurve-ADAri}
   \end{center}
\end{figure}

In Fig. 5, stellar age is plotted as a function of stellar mass.
The filled squares with error bars are look-back times and stellar
masses predicted by the De Lucia model. The look-back time of a
galaxy is the time corresponding to the redshift when 50 percent
of the stars were first formed. Open and filled circles with
arrows show galaxies that have stellar ages possibly older than
15\, Gyr (the maximum age of the BSP model). It is easy to see a
trend that more massive ellipticals have older stellar populations
and the most massive galaxies have the oldest stars. But the
changing of stellar age with stellar mass is different from the
model prediction. This is perhaps caused by the somewhat different
definitions of look-back time and stellar age. In detail, the
stellar age in the simple BSP model is defined corresponding to
the redshift when all stars formed at the same time. When we fit
the relation between stellar age and stellar mass, we find a
linear relation: age = 3.115 log(\it M$_\ast$/M$_\odot$\rm) -
24.147 , with a  0.656 correlation parameter.

%
\begin{figure}
   \vspace{2mm}
   \begin{center}
   \hspace{3mm}\psfig{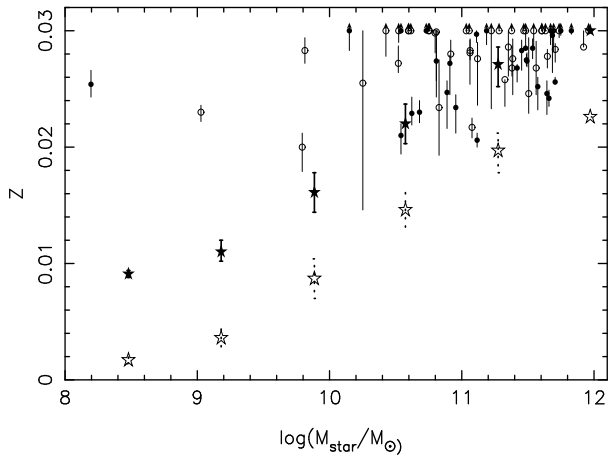}
   \parbox{180mm}{{\vspace{2mm} }}
   \caption{Stellar metallicity -- mass relation of our sample
   elliptical galaxies. Galaxies that have stellar metallicities larger than
   the maximum metallicity available for BSP model (0.03) are shown with arrows.
   The open pentacles with dashed error bars represent the predictions of De Lucia model and
   the filled pentacles with solid error bars represent the metallicities
   of De Lucia model added by 0.074. The open and filled circles have the
   same meanings as in Fig. 5.}
   \label{Fig:lightcurve-ADAri}
   \end{center}
\end{figure}
In Fig. 6, we show the stellar metallicity -- stellar mass
relation of the main sample elliptical galaxies. The open and
filled circles with arrows show ellipticals that have populations
possibly more metal rich than 0.03 (the maximum metallicity of the
BSP model). The open pentacles with dashed error bars represent
the predictions of De Lucia model. From this plot, we see that all
galaxies have metallicities richer than the predictions of De
Lucia model. However, if we only take ellipticals with stellar
metallicities lower than 0.03 into account, our data can be
expressed by a trend similar to the prediction of De Lucia model,
which can be derived by adding 0.074 (the difference between the
maximum metallicity of De Lucia model and that of the BSP model)
to each stellar metallicity predicted by De Lucia model. We plot
the trend via filled pentacles with solid error bars, which can be
seen clearly in Fig. 7. It means that our data have the same trend
as the prediction of De Lucia model. In fact, the result is
possibly limited by the theoretical models, thus the difference
between our data and prediction of the galaxy formation model is
understandable.

%
\begin{figure}
   \vspace{2mm}
   \begin{center}
   \hspace{3mm}\psfig{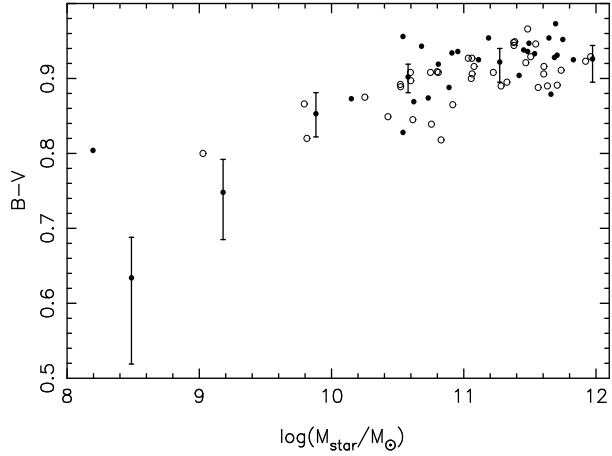}
   \parbox{180mm}{{\vspace{2mm} }}
   \caption{$B-V$ and stellar mass relation of the sample elliptical galaxies.
   The stellar masses are calculated by this work. Filled squares with error
   bars represent the values predicted by the model. Open and filled
   circles have the same meanings as in Fig. 5.}
   \label{Fig:lightcurve-ADAri}
   \end{center}
\end{figure}

The relation between galaxy color and stellar mass is plotted in
Fig. 7. The stellar masses are calculated by this work using eq.
(2). Filled circles in the plot represent ellipticals in
high-density environment and open circles represent the field
ellipticals. Filled squares with error bars represent the color
versus stellar mass relation predicted by the model. We see that
our data agree with the relation predicted by the model very well.

\subsection{Stellar age, metallicity, mass and color variation with cluster-centric distance}
The De Lucia model predicted a clear trend driven by mass
segregation and incomplete mixing of the galaxy population during
the cluster assembly. According to the prediction, within
clusters, stellar masses, ages, metal abundances and galaxy colors
would decrease on average with increasing distance from the
cluster center. To test these trends, we select 11 component
elliptical galaxies of Fornax cluster and determine their stellar
ages and metallicities from H$\beta$ and [MgFe] line indices
within r$_{e}$/8. The line indices of these galaxies are derived
from Kuntschner (2000) and their coordinates and $B-V$ colors are
derived from HyperLeda database
(http://www.brera.mi.astro.it/hypercat/). Here we show the main
data of 11 ellipticals in Table 3. It is noticeable that we use
the angular distance to the centric galaxy NGC 1399 (a galaxy well
studied, e.g. Loewenstein et al. 2005) instead of the real
cluster-centric distance for each elliptical galaxy, because it is
difficult to determine the accurate distances of galaxies while
they have uncertain peculiar velocities. The main results are show
in Figs 8, 9 and 10.

\begin{table}[]
\caption[]{Main data for 11 component elliptical galaxies of the
Fornax cluster. In the table, log$M$$_\ast$ is the logarithm of
stellar mass and $\theta$$_{1399}$ is the angular distance to NGC
1399. } \label{Tab:3}
\begin{center}\begin{tabular}{lrrrrr}
\hline\noalign{\smallskip}
\multicolumn {1} {l} {Name}& \multicolumn {1} {c}
{log($M$$_\ast$/\it M$_\odot$\rm)} &\multicolumn {1} {c}
{$\theta$$_{1399}$}&\multicolumn {1} {c} {$B-V$}&
\multicolumn {1} {c} {Age} &\multicolumn {1} {c} {Z}\\

&&\multicolumn {1} {c}{[arcmin]}&\multicolumn {1}
{c}{[mag]}&\multicolumn {1} {c}{[Gyr]}&\multicolumn {1} {c}{[Gyr]}\\

\hline\noalign{\smallskip}
NGC 1336 & 9.5886 &0.324 &0.809 &14.6       $\pm$ 4.9 &0.0179            $\pm$ 0.0029 \\
NGC 1339 &10.5695 &3.318 &0.903 &13.8       $\pm$ 2.4 &0.0270            $\pm$ 0.0030 \\
NGC 1351 &10.5559 &0.636 &0.844 &14.8       $\pm$ 2.9 &0.0235            $\pm$ 0.0039 \\
NGC 1373 & 9.1050 &0.294 &0.832 & 8.6       $\pm$ 1.9 &0.0212            $\pm$ 0.0038 \\
NGC 1374 &10.8768 &0.240 &0.894 &11.8       $\pm$ 2.3 &0.0299            $\pm$ 0.0041 \\
NGC 1379 &10.1853 &0.036 &0.866 & 9.8       $\pm$ 4.7 &0.0241            $\pm$ 0.0039 \\
NGC 1399 &12.2645 &0     &0.934 &14.1       $\pm$ 0.9 &\multicolumn{1}{l}{$\geq$0.03} \\
NGC 1404 &11.5458 &0.150 &0.941 &$\geq$15   $\pm$ 3.0 &\multicolumn{1}{l}{$\geq$0.03} \\
NGC 1419 & 9.9774 &2.160 &0.863 &14.9       $\pm$ 2.8 &0.0159            $\pm$ 0.0034 \\
NGC 1427 &10.7684 &0.084 &0.885 &11.1       $\pm$ 3.0 &0.0262            $\pm$ 0.0034 \\
IC 2006  &10.2757 &0.588 &0.896 &13.7       $\pm$ 1.0 &0.0294            $\pm$ 0.0010 \\

\noalign{\smallskip}\hline
\end{tabular}\end{center}
\end{table}

%
\begin{figure}
   \vspace{2mm}
   \begin{center}
   \hspace{3mm}\psfig{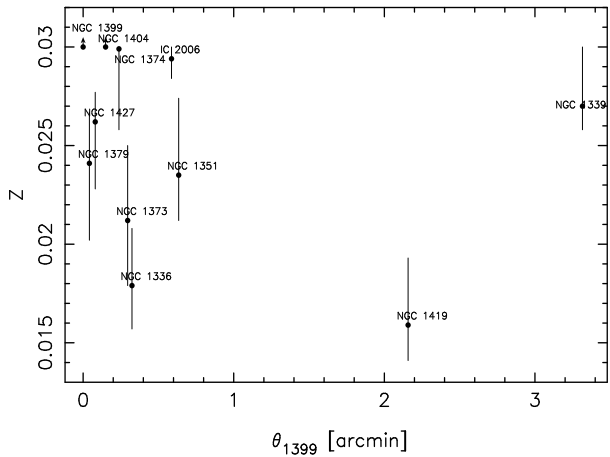}
   \parbox{180mm}{{\vspace{2mm} }}
   \caption{The plot of elliptical galaxies in (Z, $\theta$$_{1399}$) plane.
   Z and $\theta$$_{1399}$ are stellar metallicity and angular distance to
   NGC 1399, respectively. }
   \label{Fig:lightcurve-ADAri}
   \end{center}
\end{figure}

In Fig. 8, stellar metallicity is plotted as a function of angular
distance to NGC1399 ($\theta$$_{1399}$). We see that it is
difficult to find a clear trend in the whole angular distance
range. But within a small range, e.g. 2.5\, arcmin, the stellar
metallicity seems to decrease with increasing angular distance.

%
\begin{figure}
   \vspace{2mm}
   \begin{center}
   \hspace{3mm}\psfig{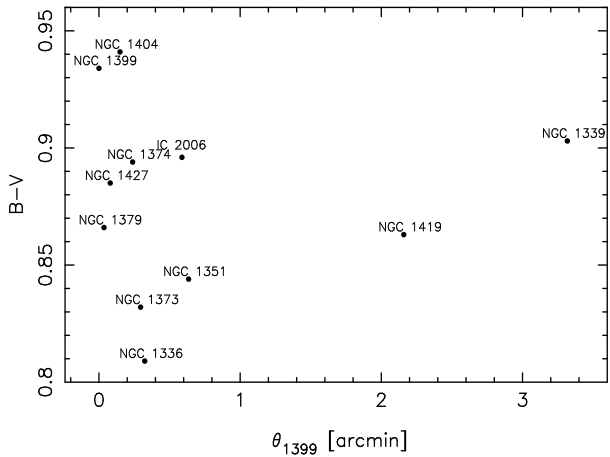}
   \parbox{180mm}{{\vspace{2mm} }}
   \caption{The plot of elliptical galaxies in ($B-V$, $\theta$$_{1399}$) plane.
   $\theta$$_{1399}$ has the same meaning as in Fig. 8.}
   \label{Fig:lightcurve-ADAri}
   \end{center}
\end{figure}
%

%
\begin{figure}
   \vspace{2mm}
   \begin{center}
   \hspace{3mm}\psfig{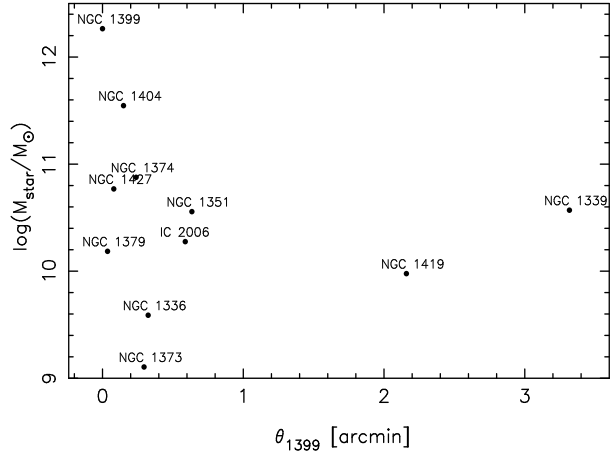}
   \parbox{180mm}{{\vspace{2mm} }}
   \caption{The plot of elliptical galaxies in [log(M$_{\rm star}$/M{$_\odot$}), $\theta$$_{1399}$] plane.
   log(M$_{\rm star}$/M{$_\odot$}) represents stellar mass of galaxies and $\theta$$_{1399}$ has the same meaning as in Fig. 8.}
   \label{Fig:lightcurve-ADAri}
   \end{center}
\end{figure}

The relation of $B-V$ color and angular distance is shown in Fig.
9 while the relation of stellar mass and angular distance in Fig.
10. The two plots do not show clear support or opposition to the
model, neither. In addition, it is found that the trend between
stellar age and mass, which we do not show here, seems almost
random.

In Figs 8, 9 and 10, we see that there are only 3 galaxies with
angular distance farther than 0.6\, arcmin, we suggest the less
elliptical galaxies that farther than 0.6\, arcmin must affect all
trends relating to cluster-centric distance. Furthermore, the
small sample of elliptical galaxies we used perhaps affects the
results.

\section{Discussion and conclusion}
\label{sect:discussion}
We determined stellar ages and metallicities of about 80
elliptical galaxies using the BSP model of Zhang et al. (2005b)
and test the latest formation model of elliptical galaxies (De
Lucia 2006) for the first time. We find that elliptical galaxies
have stellar populations about 10\, Gyr old and more metal rich
than 0.02 (see also Zhou et al. 1992).

When we analysis our data, we find that stellar populations of
elliptical galaxies in high-density environment are about 1.5\,
Gyr older while 0.001 less metal rich than those of field
elliptical galaxies. We also find that elliptical galaxies in
high-density environment are about 0.02 mag redder than field
ellipticals. Furthermore, we find that more massive ellipticals
are redder and have older and more metal rich stellar populations
than those less massive ones. It also seems that the most massive
ellipticals have the oldest and most metal rich populations.
However, elliptical galaxies in low-density environment show more
metal rich stellar populations than those high-density
complements. In fact, this trend is completely opposite to the
prediction of De Lucia et al. model. When we test the stellar
mass, age, metallicity and galaxy color variation with the
cluster-centric distance, the results do not show clear support or
opposition to the model and it seems that they are affected by
using the angular distance instead of cluster-centric distance and
the small elliptical galaxy sample we used. Therefore, the results
derived from BSP model support the $\Lambda$CDM-based hierarchical
model of elliptical galaxies formation, expect the metallicity
distribution with environments and the changing of stellar
peculiarities with cluster-centric distance. However, the
doubtless conflict between our result and prediction of the model,
i.e. the result that low-density elliptical galaxies have more
metal rich populations than high-density elliptical galaxies,
should be paid attention to.

\begin{acknowledgements}
We thank HyperLeda team for supplying us with the photometry of
galaxies on the internet: http://www.brera.mi.astro.it/hypercat/.
We also thank Prof.~Xu Zhou and Prof.~Tinggui Wang for some useful
discussions. This work is supported by the Chinese National
Science Foundation (Grant Nos 10433030, 10521001 and 10303006),
the Chinese Academy of Science (No. KJX2-SW-T06) and Yunnan
Natural Science Foundation (Grant No. 2005A0035Q).
\end{acknowledgements}

\label{lastpage}

\end{document}